# Intelligent Database Flexible Querying System by Approximate Query Processing

Oussama Tlili
Faculty of Sciences of Tunis
Campus Universitaire -1060 Tunis, Tunisia
tlili.oussama@gmail.com

Minyar Sassi
National Engineering School of Tunis
BP. 37, Le Belvédère 1002 Tunis, Tunisia
minyar.sassi@enit.rnu.tn

Habib Ounelli
Faculty of Sciences of Tunis
Campus Universitaire, Tunis 1060, Tunisia
habib.ounelli@fst.rnu.tn

*Abstract*— **Database flexible querying is an alternative to the classic one for users. The use of Formal Concepts Analysis (FCA) makes it possible to make approximate answers that those turned over by a classic DataBase Management System (DBMS). Some applications do not need exact answers. However, flexible querying can be expensive in response time. This time is more significant when the flexible querying require the calculation of aggregate functions ("Sum", "Avg", "Count", "Var" etc.). In this paper, we propose an approach which tries to solve this problem by using Approximate Query Processing (AQP).**

## I. INTRODUCTION

A flexible querying technique is used to enhance access and human interaction with information systems and to make it easier for users to find what they are looking for.

It tries to make the classic DB querying more flexible for users. To this effect, several approaches have been proposed in the literature such as additional criteria [1][2], preferences [3], distance and similarity [4][5], models based on the fuzzy-sets theory [6][7], approaches based on Type Abstraction Hierarchies (TAH) and Multi-Attributes Type Abstraction Hierarchies (MTAH) [8], and recently approaches based on the FCA [9] and those based on fuzzification of the FCA [10]. These approaches have some limits. We can mention the following:

1) No consideration of aggregate queries: they not support the aggregation functions such as *Average*, *Count*, *Max*, *Min* and *Sum*.

2) Accuracy of the answer: in many applications, the accuracy of the answer to the last decimal is not required. The user wants approached answers as soon as possible instead of waiting more time for the exact response.

3) Response time: in the case of large DB, the time taken to build the final response is enormous.

For aggregation queries, we propose a way to data route using FCA to generate a hierarchy allowing the user to personalize these responses into several levels.

For answer accuracy, we propose to use Approximate Query Processing (AQP) which consists of techniques that sacrifice accuracy to improve response time.

To improve response time, we propose to adapt the online aggregation [11] whose objective is to gradually approximate answers when running the application. It consists of applying a sample on the initial data of the DB to minimize disk access and therefore improve response time.

This paper is organized as follows. After the introduction, Section 2 presents a state of the art on flexible querying systems recently proposed and techniques of AQP. In Section 3, we propose the architecture of our system. In Section 4, we detail the various steps of the proposed approach. In Section 5, we present a general description of our approach by an illustrative example. In Section 6, we make a comparative study between the proposed approach and approaches similar to ours. In Section 7, we evaluate our approach. Finally, we summarize our work and propose future works in Section 8.

## II. STATE OF THE ART

In this section, we present flexible querying systems and some AQP techniques.

### A. Flexible Querying Systems

Flexible querying database try to extend the binary querying by introducing preferences in query criteria. These preferences allow for direct qualitative responses. Thus, data returned by a query will be "more or less relevant", according to the preferences.

Research on flexible querying investigates the handling of imperfectness of information (about queries), e.g., due to imprecision, uncertainty and/or incompleteness. Using traditional querying techniques, a record will only be part of the query result if it completely satisfies all the constraints imposed by the query. Due to imperfections, which often occur in reality, such an approach is too stringent. Also, in traditional querying a query is generally a complete specification of what is wanted. Flexible querying helps to relax this, making it possible that records that e.g., satisfy most (but not all) of the constraints will also be present in the query result –this is particularly useful when none of the records satisfies all constraints– and allowing query formulations to be invariably incomplete.

In this section, we limit ourselves to the approaches close to our.

Query relaxation approach proposed in [8] uses predicates with relaxing attributes. In this context, we use attributes with predicate for comparison with a linguistic term such as "Average" in place to say "Price between 200 and 300".

This approach present two main contributions compared to others especially that of Chu *et al.* [12]. These contributions are as follows:
- Taking into account the interdependence of the search criteria query.
- Detection of inconsistencies between the search criteria before executing the query.
- Cooperation with the user by offering data near the query instead of empty answers.

However, problems of storage and indexing HAT and HATM structures constitute a handicap to their use in the querying process.

In [10], fuzzification of the FCA in the process of flexible querying was introduced. The general principle of this approach is to organize data to optimize the query towards his given. The notion of concept application is used to allow verification of the query realisability. The returned answers were classified by satisfaction degree measured compared to the user query.

Some limits arise with this approach. We can mention: i) the response time met for answers generation, and ii) the complexity of the used structures.

A cooperative approach to flexible relational DB querying proposed in [9] based on fuzzy set theory to model the fuzzy predicates included in the query. It is based on the lattice concept to evaluate flexible queries submitted by users.

Moreover, the approach generates query causes with no answers and offers sub-queries with approximate answers. However, the approach has several limitations such as:
- Scheduling of sub-queries approximate taking into account preferences expressed by the user in the original complaint.
- The inclusion of some widely used language modifiers like "most" and "approximately" in the query qualifiers.

All these approaches do not take into account agregate queries and have a response time sufficiently high.

### B. Approximate Queries Processing by sampling

The AQP is an effective solution which consists of techniques that sacrifice accuracy to improve response time. It is used in aggregate queries (including SUM, COUNT, AVG, etc ...), whose accuracy what the "last decimal" is not required.

There are several techniques for the AQP, we can cite the sampling techniques [13], the use of histograms and Wavelets [14].

We are interested in sampling techniques. His principle is to build tables or views by selecting certain rows from the table to build an initial sample. It has a storage size smaller than the initial table, instead of questioning all the comics, the user asks a sample representing the DB and then gets an approximate answer.

The basic architecture of AQP based on sampling as described in Figure 1. It consists of two phases:
- **Offline Phase**: before executing the query, the sample is constructed from the DB tables.
- **Online Phase**: queries are rewritten to be run on the sample. The result is then measured to give the approximate response also with an error rate.

### III. ARCHITECTURE OF THE PROPOSED SYSTEM

Figure 1 describes the querying flexible system architecture called FLEXTRA. We have added several components to relational DBMS such as KB (Knowledge Base).

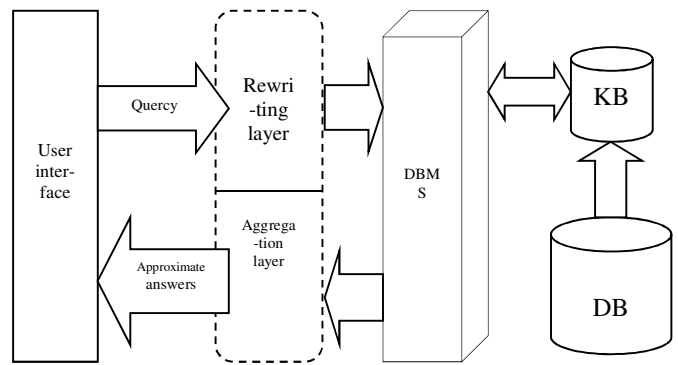

Figure 1. System Architecture

his system includes the following components:
- Rewritable layer: it takes care of rewriting the aggregate query in its final form by adding aggregate functions and calculating the error rate depending on the confidence degree defined by user. The query becomes an approximate query.
- Aggregation layer: it is responsible for transferring the user with different responses gradually during the query execution. It gives the error rate.
- DB: it is a relational database where we store all permanent information in a relational model.
- KB: it is a Knowledge Base that is generated from the DB and before the query execution. It contains information on the relaxing attributes (an attribute that describes a linguistic term). The schema is described in Table 1.

TABLE 1. KB SCHEMA

| ID row | Relaxing-Attribute1 | Relaxing-Attribute 2 | ... | Relaxing-Attribute n |
|---|---|---|---|---|
| ... | ..... | ..... | ..... | ..... |

### IV. DESCRIPTION OF THE PROPOSED APPROACH

Our approach is described in Figure 2. It is divided into two major phases:
- Pre-treatment phase: in this phase we will generate the KB from the DB to contain the degrees of membership of each tuple relaxing attributes.
- Post-treatment phase: when the user launches the application, the system searches for approximate answers,

and then calculates the aggregation and gradually sends to the user.

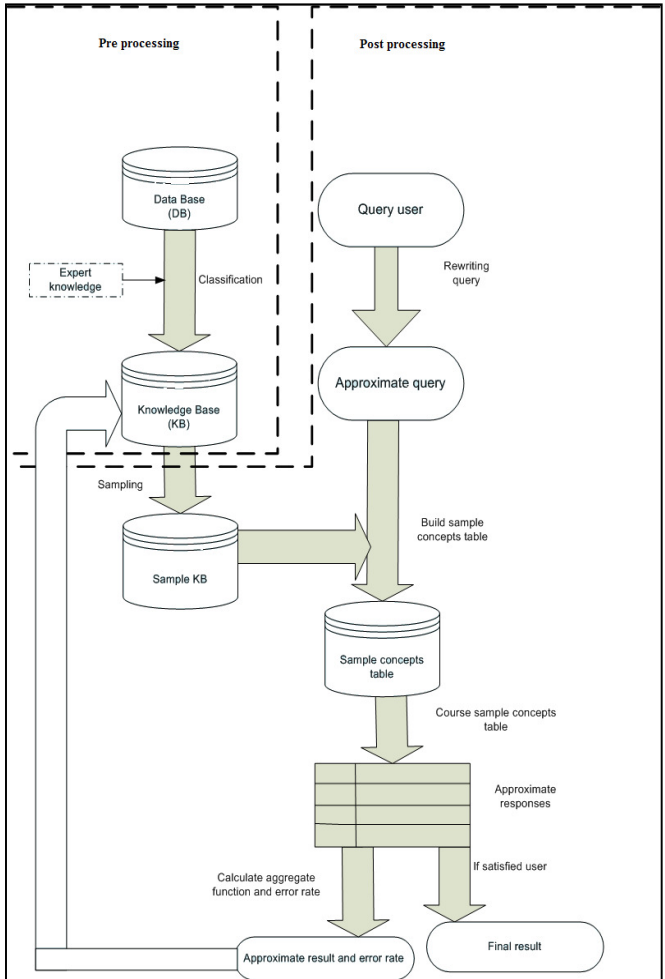

Figure 2. The approach phases

### A. Building KB

Clustering allows partitioning the data into clusters, the domain expert will assign linguistic terms (e.g. young age, low salary…) to use it in the query and this constitutes the KB.

A KB contains the membership degree of each tuple to relaxing attributes using the membership function. Zadeh proposes a series of membership functions [15], we include essentially the triangular function, the function singleton, L Function, Gamma function, and trapezoidal function.

We use a trapezoidal function, it is defined by a lower limit $a$, an upper limit $d$. Moreover, it is characterized by a lower limit $b$ and an upper limit $c$ to the core. This function is defined as follows:

$$\mu_E(x) = \begin{cases} 1 & si\ c \leq x \leq b \\ \dfrac{x-a}{b-a} & si\ a \leq x \leq b \\ \dfrac{d-x}{d-c} & si\ c \leq x \leq d \\ 0 & si\ x < a\ ou\ x > d \end{cases}$$

**Example:** Table 2 presents the membership matrix on "Age" attribute; it has two relievable attributes "young" and "adult".

TABLE 2. FUZZY CLUSTERING IN AGE ATTRIBUTE

| Id row | Young Age | Adult Age |
|---|---|---|
| 25 | 0.7 | 0.3 |
| 30 | 0.2 | 0.8 |
| 20 | 1 | 0 |

If Age_Young= 0.7 to 1 then the row has a membership degree = 0.7 for the Young_Age cluster.

### B. Query Flexible Rewriting

The first step of query execution consists of construct the approximate query through an interface in which it specifies the confidence degree, the target table, the aggregate function (SUM, AVG, COUNT ...), all attributes of the SELECT clause and all attributes of the WHERE clause.

In this paper, we consider viewing a single table without using Group By knowing that it contains thousands of records. The approximate query as follows:

*Select function(attribute), confiance_degree as confidenc, functionInterval(confiance_degree) from table where attribuet1 IS flexible_ condition1 [and … attribute2 IS flexible_ condition2]*

Where *function( )* and *functionInterval( )* [11] are user predefined functions and which can give online approximate answers depending on the confidence degree for aggregate *AVG, SUM, COUNT ….*

### C. Sampling

The second step of our approach is to construct a sample from the KB.

The sampling is made in online mode and the gain of time is due of treatment of small KB (not all KB where construction of context table in large KB takes a long time).

Instead of querying the entire KB, we interview a sample of KB that is made up of hundreds of records which improves response time.

Administrator (expert) sets the percentage of sampling depending on the size of KB (if $s$: Percentage of sampling and $n$ = the size of KB then sample size $p = (n * s)/100$.

I use the method of [11] in sampling, I choose randomly $p$ lines from KB who have not previously extracted.

Our approach is to build the sample using the following algorithm:

```
Algorithm1 : Sampling
```
**Inputs:** Query :Q
       Knowledge base :KB
       KB size :P
       Sample Percentage: p
**Output** :Sample :E
**Begin**
**Step 1**: KB1:= KB-E.
**Step 2**: E contains the $\frac{n*p}{100}$ lines extracted randomly from KB1.
**Step 3**: **Repeat** steps 1 and 2 **Until** all rows have been processed.
**End**

### D. Building sample concepts table

The third step of our approach is to build the concepts table associated with sample building in the previous phase. The concepts table [16] is a tabular representation of a concept lattice and its construction is easier than the lattice.

The context table is a table structure but not a tree (concept lattice) and it is simple to use, modify, delete and generate concepts in the implementation step.

The context table is simply the result of a clustering operation giving membership degrees of each data to each cluster.

This is described in Table 3, where the columns have the following meanings:
- C# (context#): The name of the source context.
- Niv#, N#(Level#, Node#) :These two columns store the identifier of the concept of context. The first is the level of the concept in the lattice while the second represents the sequence number of the concept at this level.
- Int#, Ext# (Intention, Extension) These columns store for (respectively extension) of each concept.
- L_s#,L_p#(Successors List, predecessors list): These two columns store the identifiers of successors (predecessors respectively) of the concept..
- T_i,T_e (Size_Intension, Size_Extension): These two columns store the cardinality of a concept (respectively the number of attributes and the number of objects).

TABLE 3. SAMPLE CONCEPTS TABLE

| C# | Niv# | N# | Int# | Ext# | L_s# | L_p# | T_i | T_e |
|----|------|-----|------|------|------|------|-----|-----|
| ….. | …. | … | … | … | …. | …. | .. | …. |

### E. Coursing the sample concepts table and calculating agregation

In this step, we course the sample concepts table to extract approximate answers and to calculate the final result of approximate aggregation.

We use algorithm proposed in [16] to build a sample concepts table on the approximate query and then return the approximate answers. In order to improve the response time, we build the concepts table using only the query conditions.

This reduces the table size and minimizes the complexity of the construction of the sample concepts table.

We calculate the aggregation function (AVG, SUM and COUNT), using the algorithm 2 with the following descriptions:
- value (t): represents the aggregate value of the tuple *t*.
- degree (t): represents the membership degree of *t*.

To calculate the aggregation, we use these functions:
- For *AVG( )* function :
$$AVG = \left(\frac{1}{n}\right) degree * \sum_{i=1}^{n} v(Li) \quad (1)$$
- For *SUM( )* function:
$$SUM = degree * \sum_{i=1}^{n} v(Li) \quad (2)$$
- For *COUNT( )* function:
$$COUNT = degree * \sum_{i=1}^{n} 1 \quad (3)$$

Where $degree = Min(U_{i1} \wedge V_{i1} \wedge .... \wedge Z_{i1})$, and $U, V, Z$ are the membership degrees on the query $Q$ and $n$ is the sample size, $v(Li)$ is the value of the tuple index $i$ ($L_i$ is a random index).

We calculate the error rate (Interval) associated with the aggregate function. We use the method of conservative confidence intervals [11]:

$$Error\ Rate = (b-a)\left(\frac{1}{2n} ln\left(\frac{2}{1-p}\right)\right)^{1/2} \quad (4)$$

Where [*a, b*] is a predetermined interval, such that $a \leq v(i) \leq b$ for all $1 \leq i \leq m$, $n$ = sample size, $m$ = size of KB, $p$ is the setting of confidence (example $p = 0.95$).

```
Algorithm 2 : Calculate_function
```
**Inputs**: concepts table: TCX
      Maximum value of attribute :max
      Minimum value of attribue : min
      Sample size :n
      Aggregate function: f
**Outputs** : result : res, Error rate :rate
**Begin**
  D=1
  som=0
  card=0
  **For each** element E of the concept table TCX
    **if** extension $\neq \emptyset$ **then**
      **for each** objet t of the extension
        som=som+value(t)
        card=card+1
        **if** degre(t)< D **then**
          D=degre(t)
        **End if**
      **End for**
    **End if**
  **End for**
  **If** f= avg **then**
    res=(som/Card)*D
  **else if** f=sum **then**
    res=som*D
  **else if** f=count **then**
    res=card*D
  **end if**
  rate= $\frac{1,22*(max-min)}{\sqrt{n}}$
**End**

## V. ILLUSTRATIVE EXAMPLE

Let a simple relational table *"employee" (id, name, age, salary)*, which contains the following rows (see Table 4).

TABLE 4: EXAMPLE OF THE RELATIONNAL TABLE EMPLOYEE

| ID | Name | Age | Salary |
|---|---|---|---|
| 1 | MOHAMED | 23 | 400 |
| 2 | ALI | 30 | 550 |
| 3 | WALID | 45 | 700 |
| …… | …… | ……. | …….. |
| 10000 | WAJDI | 40 | 800 |

The relaxing attributes *Age-Young, Age-Adult, age-Low, Salary-Middle, Salary-High*, and KB which contains rows as shown in Table 5:

TABLE 5 : CLUSTERING DATA OF THE RELATION EMPLOYEE

| ID tuple | Age-Young | Age-Adult | Salary-Low | Salary-Middle | Salary-High |
|---|---|---|---|---|---|
| 1 | 0.7 | 0.3 | 0.6 | 0.4 | 0 |
| 2 | 0.5 | 0.5 | 0 | 1 | 0 |
| 3 | 0 | 1 | 0.1 | 0.6 | 0.3 |
| … | …. | …. | ….. | …. | …. |
| 10000 | 0.1 | 0.9 | 0 | 0.3 | 0.7 |

Then, we eliminate data with low membership degree by setting a user defined threshold, KB becomes as shown in Table 6:

TABLE 6 : CLUSTERING DATA OF THE RELATION EMPLOYÉE WITH A THRESHOLD

| ID tuple | Age-Young | Age-Adult | Salary-Low | Salary-Middle | Salary-High |
|---|---|---|---|---|---|
| 1 | 0.7 | - | 0.6 | 0.4 | - |
| 2 | 0.5 | 0.5 | - | 1 | - |
| 3 | - | 1 | - | 0.6 | - |
| … | …. | …. | ….. | …. | …. |
| 10000 | - | 0.9 | - | - | 0.7 |

Consider the following query for finding the average salary for young employees and low salary with a confidence level = 95%.

> *"Average Salary of Young employees and Low Salary with a degree of confidence= 95% "*
>
> *"Select Avg(Salary) from employees where age IS Young and Salary IS Low"*

The approximate query becomes:

> *"Select AVG (Salary), 0.95 as confidence, ConsAvgInterval( 0.95) from employee where age IS Young and Salary IS Low"*

We construct the sample (Table7) according to the KB at the time of query execution.

TABLE 7. SAMPLE OF DATA

| ID row | Age-Young | Salary-Low | Salary |
|---|---|---|---|
| 1 | - | 1 | 400 |
| 20 | 0.8 | - | 900 |
| 520 | - | 0.9 | 430 |
| 32 | - | 0.8 | 460 |
| 10 | 0.6 | - | 780 |
| …… | …… | …… | …… |
| 130 | - | 0.5 | 550 |

Then we generate a concepts table associated with the query as shown in Table 8.
With each given extension contains two attributes: The first is the degree and the second is the aggregated value.
**Example**: 20 (1, 380) the row 20, a degree is 1and its value is 380.
We repeat these steps until all the KB is treated either we get an error rate is very low to say the exact result is very close to either the user is satisfied with the outcome and conclusion the query execution.

TABLE 8 : SAMPLE CONCEPTS TABLE

| C# | Niv # | N# | int# | Ext# | L_# | L_p# |
|---|---|---|---|---|---|---|
| 1 | 1 | 1 | Young_A low_S | ∅ | (1,2,1) (1,2,2) | 0 |
| 1 | 2 | 1 | low_S | 1(1 ;400) 32(0,8 ;460) 520(0,9;430) 130(0,5;550) | (1,3,1) | (1,1,1) |
| 1 | 2 | 2 | Young_A | 10(0,6 ;780) 20(0,8;900) | (1,3,1) | (1,1,1) |
| 1 | 3 | 1 | ∅ | 1(1 ;400), 32(0,8;460), 520(0,9;430) ,130(0,5;550) 10(0,6 ;780), 20(0,8;900) | 0 | (1,2,1) (1,2,2) |

In Table 9, we present an example of results returned after the calculation AVG and error rate functions.

TABLE 9: RESULTS OF APPROXIMATE ANSWERS

| AVG | Confidence | Error rate |
|---|---|---|
| 400 | 95 % | 0.06503 |
| 402 | 95% | 0.06500 |
| 405 | 95% | 0.06470 |
| …. | … | ……. |
| 410 | 95% | 0.0090 |

## VI. COMPARATIVE STUDY

In this section, we present the essential idea of the main approaches to flexible querying the closest to ours. We specify each time different art studies conducted on these approaches. They differ mainly by the way used to find the values closest to those requested by the user and the formalism used to model uncertainty and imperfection of the real world.

The contributions of the approach Ounelli *et al*. [8] are important, including the TAH and MTAH concepts for modeling generalization and specialization hierarchies of concepts. In this approach, no modification of SQL is required, what constitutes an asset for the implementation of this approach. The user does not apply during the relaxation to make choices that can be hazardous.

In this approach, the relaxing attributes are set by the administrator of the DB. This is especially important that the proposed approach is aimed at end users with no specific and detailed knowledge on the organization and the data they consult. It is easier for an expert to specify a *price* attribute of the *DB* table is relaxing and can be used with the terms *"low", "comfortable"* or *"high"*.

However, this approach has limitations in the structures it uses. We mainly include: i) incremental maintenance of the KB relaxing attributes, ii) clustering of relaxing attributes without fixing a priori the number of clusters, iii) the problem of storage ,clustering and indexing HATM, and iv) not taking into account aggregate queries.

In the approach of Sassi *et al*. [10], generated clusters for each relaxing attribute are not stored in the DBMS catalog. Thus, the maintainability of this meta-base is no longer a problem. Indeed, in order to draw the concept lattice, core of FCA, they must simply load an XML file that can retrieve all the information necessary to trace these lattices.

However, this approach has limitations in the structures they use. We mainly include i) the number of concepts generated, ii) the response time used to generate approximate answers, and iii) not taking into account aggregate queries.

The approach of Chettaoui *et al*. [9] allows the treatment of empty response to a flexible query. Thus, it detects the causes of failure and allows the generation of sub-queries and approximate answers.

Another advantage of this approach is that not changing the structure of SQL and thus benefit from the features of the DBMS.

However, this approach does not allow the use of linguistic modifiers in the query. This test is interesting since users typically use such linguistic terms and it does not take into account the aggregate queries.

The approach of Hass *et al*. [11] allows classical querying (Boolean) on broad comic returning relevant answers in the shortest time for aggregate queries. It aims to gradually give approximate results when the query execution until all data has been processed. Thus, the user observes the degree of progress of the response and controls the query execution. We are not obliged to wait several minutes for the query result.

The approach proposed in this paper combines the advantages of those mentioned above while overcoming the limitations they present.

Indeed, we can perform an aggregate query on large flexible DB while returning relevant answers in the short time and the error rate.

Table 9 presents a comparative study between the approaches mentioned above and ours.

TABLE 9. COMPARISON BETWEEN DIFFERENT APPROACHES

|  | Agregation | Sampling | Flexibility | Accuracy |
|---|---|---|---|---|
| Query Relaxation | - | - | X | - |
| Fuzzification of concepts lattice | - | - | X | - |
| Online AQP | X | X | - | X |
| Flexible interrogation by AQP | X | X | X | X |

## VII. EVALUATION

Figure 3 shows the main interface of the FLEXTRA system.

The responses appear in the table when executing the query. In this case, the user does not have to wait until runtime to have the final result. Indeed, after a while, the initial response and the error rate is displayed, until the user stops the calculation or that KB has been completely treated.

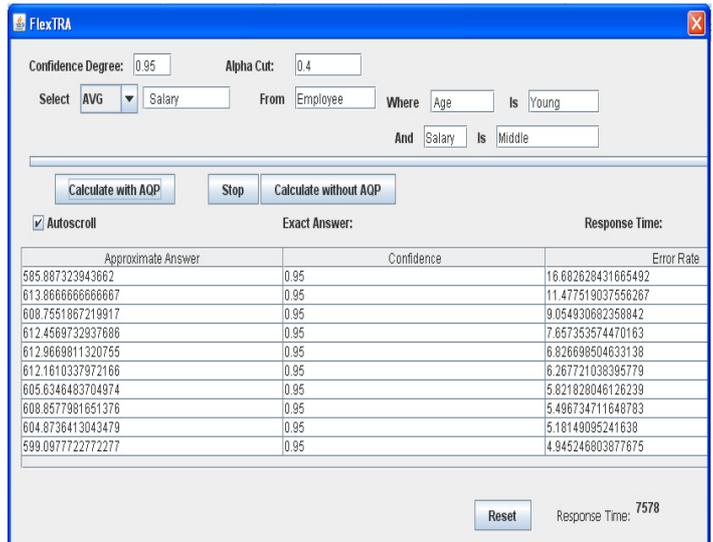

Figure 3. Display approximate answers

### A. Testing the response time

We started with the table employee (id, name, salary, age) and we increased the number of records from 789 to 9498 records and calculate for each case the response time of FLEXTRA system and compare it with the case of AQ (approximate query) and classic querying(without AQ), as shown in Figure 4.

For the employee table, it contains two relaxing attributes *"age"* and *"salary"*.

The query is: "the average low salary of young employees"

*"SELECT AVG(salary) FROM employee WHERE salary IS low AND age IS young "*.

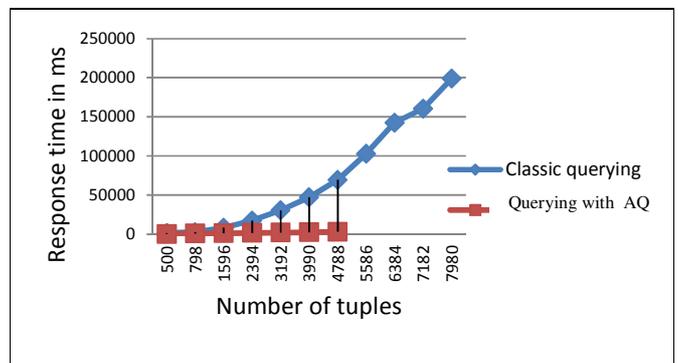

Figure 4. Comparison between the two approaches

From Figure 4, we find that the response time is lower using the AQ than classing querying.

For the classical query, the curve is exponential, while for interrogation with AQ, the curve is linear.

If the size of the database exceeds 7000 records, the response time for classic querying, is about 2 minutes, so it is with the AQ, the order of 5 seconds.

### B. Testing the accuracy response

We will run the application in the table "employee" in both cases: with and without AQ, then check the quality of answers returned with AQ.

For the table employee, the exact answer is 34.9 years to have when we execute the following query:

"SELECT avg(age) FROM employee WHERE age IS adulte AND salary IS middle "

Figure 5 shows the development of approximate answers to reach the exact answer:

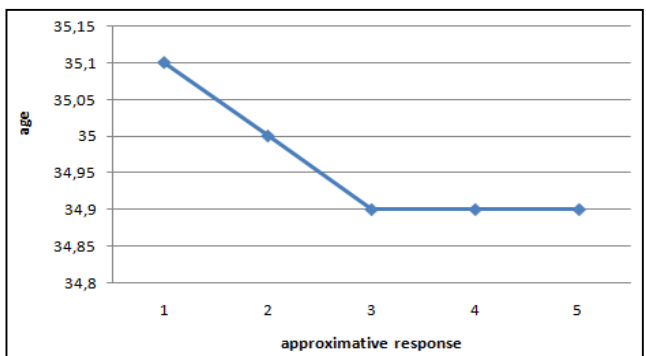

Figure 5. Evolution of approximate answers

We note that approximate answers are near to the exact answer, the first approximate answer is 35.1 years + / - 1.2 with 95% confidence level, whereas the correct answer (classic querying) is 34.9 years.

## VIII. CONCLUSION

We proposed in this work a flexible querying approach of large DB while using AQ.

It is about a field in strong expansion. On the one hand, the DB are increasingly bulky. In addition, the construction of querying systems able to satisfy flexible queries is more complex and expensive.

We integrated the AQ techniques in a system using FCA in order to overcome limits of the existing approaches when we use aggregate queries ("Sum", "Avg", "Count", "Var" etc.) such as the response time and confidence rate of result answers.

The type of query is "give the average of the weak salary"

This system makes it possible to turn over quickly approximate answers while holding trying to improve the exactitude of the provided answers.

Our approach comprises two steps:

- Pre processing step in which the KB is generated starting from the DB so that it contains membership degrees of each tuple to the relievable attributes.
- Post processing step during which the flexible query is rewritten so that it becomes an AQ. The sampling of the KB consists in extracting some data (tuples). The construction of a sample concepts table is made to release the approximate answer and the Error Rate.

For the exactitude of the answer, we used AQ which support the response time to the detriment of the result exactitude.

In order to improve the response time, we propose in this article to adapt online aggregation technique proposed in [11], whose objective is to gradually give approximate answers while executing the query.

It consists to apply a sampling to the initial data of the DB in order to minimize the disc access and consequently to improve the response time.

Our approach contributes several shares in particular:
- The calculation of aggregation for flexible queries.
- The improvement of response time by guaranteeing the exactitude of the answer.
- The processing of the case of empty answers for a flexible query.
- No modification of the structure of the DBMS and SQL language.
- The use query execution control.

To implement this approach, layers will be added to a conventional DBMS such as:
- Rewritable layer: it takes care of rewriting the query aggregation for which an application becomes rough.
- Aggregation layer: it is responsible for calculating the responses gradually during the query execution.

As futures works, we propose:
- The integration of complex and nested queries involving join operation.
- The calculation of the aggregation functions on several attributes.
- The inclusion of some widely used language modifiers like "very" and "approximately" in the query qualification.
- The use of other sampling procedures in order to improve the confidence rate.